\documentstyle[psbox,psfig]{workshop}

\def\etal{{et\,al. }}

\def\grad{$^\circ$}
\def\dashdot{\cdots\cdots\cdots\cdots\cdots\cdots\cdots\cdots\cdots\cdots
\cdots\cdots\cdots\cdots\cdots\cdots\cdots\cdots}

\def\exo{{\sl EXOSAT }}
\def\ein{{\sl Einstein }}

\def\amin{\ifmmode ^{\prime}\else$^{\prime}$\fi}
\def\asec{\ifmmode ^{\prime\prime}\else$^{\prime\prime}$\fi}
\def\degs{\ifmmode ^{\circ}\else$^{\circ}$\fi}
\def\fdg{\hbox{$.\!\!^\circ$}}          % Fractions of degrees

\begin{document}

\title{Gamma-Ray Bursts with ROSAT}

\author{Jochen Greiner\thanks{Present address: Astrophysikalisches Institut 
Potsdam, 14482 Potsdam, Germany; email: jgreiner@aip.de} 
\\[12pt]
Max-Planck-Institut f\"ur extraterrestrische Physik, 85740 Garching, Germany}

\abst{I review the use of ROSAT over the last years for the investigation of
well localized gamma-ray burst (GRB) error boxes. In particular, I cover 
(i)  the systematic study of several dozens of IPN locations using the
ROSAT All-Sky-Survey data, 
(ii) results of deep ROSAT pointings of selected small GRB error boxes,
(iii) the attempts for and results of quick
follow-up observations after GRB events including the three GRBs localized
with BeppoSAX,
(iv) the correlation of GRB locations with serendipitous ROSAT pointings and
(v) the search for X-ray flashes in the database of pointed ROSAT observations.
}

\kword{Gamma-rays: bursts --- X-rays: general --- satellite: ROSAT}

\maketitle

\section{Introduction}

The emission of $\gamma$-ray bursts (GRB)  has been measured over a wide
range of photon energies ($\approx$2 keV--2 GeV). 
While most of the emission is at  0.1--10 MeV 
the fraction of X-rays to the total  power is typically
less than a few percent (Higdon \& Lingenfelter 1990) during the bursts. 

No reliable quiescent counterparts of GRBs have been detected at other 
wavelengths, and the basic  pre-ROSAT results on X-ray counterpart searches
can be summarized as follows:
(1) \ein observations of 5 GRB error boxes (Pizzichini \etal 1986) have 
revealed one possible point-like source in the GB 781119 error box,
(2) \exo observations
resulted in upper limit estimates of 5$\times$10$^{-13}$ erg/s/cm$^2$ 
for the 3 error boxes of GRB 790305B, 790418 and 791105B  (Bo\"er \etal
1988, 1991), (3)
\ein HRI images of the SNR N49 show enhanced X-ray emission coincident with 
the error box of GRB 790305B, the famous  Soft Gamma Repeater, 
but identification is still unclear (Rothschild \etal 1994).

These results have motivated extensive searches with the more sensitive ROSAT 
satellite.
While the above results have been obtained from dedicated pointed observations,
ROSAT offers a few additional possibilities with respect to GRB research which 
will be reviewed here for the first time. References to individual ROSAT 
results already published in the literature will be given in the corresponding 
subsections below.

\section{ROSAT All-Sky-Survey and GRB error boxes}

\subsection{ROSAT All-Sky-Survey and Data Analysis}

Between 1990 Aug. 1 and 1991 Jan. 25 the ROSAT satel\-lite (Tr\"umper 1983) 
has performed the first all-sky survey in the soft (0.1--2.4 keV) X-ray energy 
band (with short additions in 1991 Feb. 16--18 and Aug. 4--12, as well as
1997 Feb.). This dataset offers the unique possibility to investigate
the soft X-ray content of a large number of GRB error boxes in a homogeneous
approach. 

Such an investigation has been performed by selecting  the $\approx$50 smallest 
GRB error boxes, in particular 36 from the 2nd interplanetary network (IPN)
1978--1981 (Atteia \etal 1987), 15 from the 3rd IPN 1991--1993 (Hurley 
1993) and 3 COMPTEL/Ulysses error boxes and the three Soft Gamma 
Repeaters (SGRs).
This means, that ROSAT data are taken either about 10 years AFTER the bursts
(2nd IPN), or a few months to years BEFORE the bursts.
Due to the scanning scheme during the all-sky survey observation the exposure 
time varies across the sky, i.e. increases towards higher ecliptic latitude. 
For the GRB error boxes investigated here, the range of exposure times is 
70 -- 1350 sec.

The data analysis involves several distinct steps and is described in more
detail elsewhere (Greiner \etal 1995a). In short, first, a 
source detection in the three adjacent energy bands 0.1--0.4 keV,
0.5--0.9 keV and 0.9--2.4 keV is performed to find all X-ray sources with a 
likelihood of existence $>$8 (corresponding to about 3.5 $\sigma$); 
second, upper limits of the countrate and the likelihood are calculated (again
in the three adjacent energy bands)  for every spatial resolution element 
(thus resulting in upper limit countrate maps and likelihood maps in each of 
the three energy bands)
to help assessing the reality of 
low-significance fluctuations (below 3.5 $\sigma$);
%\item Division into three energy bands gives valuable information on the
%general spectral shape, and hence on the possible nature of the object.
and third, optical follow-up work (mainly spectroscopic observations)
has been performed for the identification of X-ray sources.
The analysis has been extended to the nearby surrounding of each box (adding 
half of the GRB error box widths on each side) which is used as control area.

\subsection{Results}

Excluding the SGRs in the following, the basic results can be summarized as 
follows (preliminary results have been reported already in Greiner \etal 1991, 
Boer \etal 1994 and Greiner \etal 1995a):

\begin{enumerate}
\vspace{-0.3cm}
\item We found X-ray sources in only five of the selected GRB error boxes.
All five sources have bright active stars (coronal X-ray emission)
as optical counterparts.
\vspace{-0.3cm}
\item The total area of all selected error boxes is 3.5\,$\Box$\grad.
With a mean density of ROSAT all-sky survey sources of 1.5/$\Box$\grad\,
one expects 5.3 non-correlated X-ray sources as compared to the 5 sources
detected.
This statistically supports the conviction that the above mentioned coronal
X-ray sources are not related to the GRB sources.
\vspace{-0.3cm}
\item The control area  is 3$\times$  larger than
the GRB boxes themselves, i.e. 10.5 $\Box$\grad, thus implying
$\approx$16 non-correlated X-ray sources. In total, 17 X-ray sources are
detected, and most of them have ``ordinary" optical counterparts.
\vspace{-0.3cm}
\item There exists no systematic correlation between GRB error boxes and 
optically unidentified soft X-ray sources (basically all X-ray sources are
optically identified).
\vspace{-0.3cm}
\item Even at lower likelihood values of X-ray source detection (i.e. lower
source intensity) there is no trend of clustering in/near error boxes.
\vspace{-0.3cm}
\item Lasota (1992) has argued that a burst will produce a wind, supersonic 
expansion of which would lead to a strong shock. Thus, accretion would cease
for several years after the burst. Accordingly, the ``null" result 
concerning the 2nd IPN error boxes might be expected. But since we also find 
no correlation for the 3rd IPN GRB error boxes, the above scenario probably 
does not apply to GRBs (see more extensive discussion in Greiner \etal 1995b).
%\vspace{-0.3cm}
\end{enumerate}

\section{Deep ROSAT observations of GRB error boxes}

Over the last six years nearly a dozen GRB error boxes were observed with the
ROSAT satellite for 10--40 ksec exposure time (see Tab. \ref{deep} for a 
complete listing), thus improving considerably the sensitivity limits obtained
with earlier X-ray observations at soft energies. For a part of these GRB 
locations, X-ray sources have been found inside the error boxes. Most notably
these include GRB 781119 (the GRB with the earlier Einstein detection) and
GRB 920501, the X-ray sources of both of which have been also detected in
ASCA observations. In the case of GRB 781119 even a second X-ray source
has been found inside the error box. 

While originally the discovery of
a quiescent X-ray source inside a small GRB error box has been considered as
probable evidence for an association of a GRB with a quiescent counterpart,
the continuing discovery of further X-ray sources and in particular the
detection of more than one X-ray source even in small GRB error boxes 
(see Tab. \ref{deep}) makes this association doubtful. Also, the optical
identification of these X-ray sources, though not yet completely established 
in all cases, does not find evidence for unusual objects.

An estimate of the chance probability for the occurrence of a quiescent
X-ray source inside a well-localized GRB error box depends on how the 
question is asked in detail (see Hurley \etal 1996 for various possibilities).

   \begin{table}[h]
     \caption{Dedicated ROSAT pointings on GRB locations (non-TOO)}
     \begin{tabular}{crcccl}
      \noalign{\smallskip}
      \hline
      \noalign{\smallskip}
 GRB$^{a)}$ & $\!\!$Exposure$\!\!$ & Date & Inst. & N$_{\rm X}^{b)}$ & Ref.$^{c)}$ \\
            &  (sec)~~             &      &  $\!$P/H$^{d)}$$\!$    &  &     \\
       \noalign{\smallskip}
      \hline
      \noalign{\smallskip}
       781119  &  2513~~ & 950110 & H & 2 & 1 \\
               & 41164~~ & 951221 & H & 2 & 1 \\
       790313  & 25111~~ & 950209 & H & 3 & 2, 11 \\
       790325B & 14649~~ & 910404 & P & 0 & 2 \\
               &   180~~ & 910416 & P & 0 & 2 \\
       790406  & 19018~~ & 911204 & P & 0 & 3, 11  \\
       790613  & 13880~~ & 910320 & H & 1 & 4, 11 \\
       881024  & 10978~~ & 910410 & P & 3 & 4, 11 \\
       910219  & 14422~~ & 931105 & P & 0 & 2, 11 \\
       910601  & 20807~~ & 921112 & P & 0 & 5 \\
       910814  & 25254~~ & 920616 & P & 1 & 6  \\
       920501  & 30386~~ & 931023 & P & 1 & 7 \\
               &  6215~~ & 940507 & H & 1 & 7 \\
       920525  &  9713~~ & 941002 & H & 0 & 6  \\
      \noalign{$\dashdot$}
      790305 & 20999~~ & 920317 & H & 1 & 8 \\
             &  9529~~ & 951010 & H & 1 & 9 \\
             &  4280~~ & 951109 & H & 1 & 9 \\
             & 12472~~ & 951209 & H & 1 & 9 \\
             &  7721~~ & 960111 & H & 1 & 9 \\
             & 12390~~ & 960207 & H & 1 & 9 \\
             &  9578~~ & 960308 & H & 1 & 9 \\
             & 10978~~ & 960416 & H & 1 & 9 \\
             & 10890~~ & 960516 & H & 1 & 9 \\
             & 15313~~ & 960615 & H & 1 & 9 \\
             & 11428~~ & 960716 & H & 1 & 9 \\
             & 15675~~ & 960814 & H & 1 & 9 \\
             & 10794~~ & 960913 & H & 1 & 9 \\
    1806--20 &  2985~~ & 910318 & P & 0 & 10 \\
             & 10078~~ & 930402 & P & 1 & 10 \\
      \noalign{\smallskip}
      \hline
      \noalign{\smallskip}
   \end{tabular}
   \label{deep}

   \noindent{\small 
       $^{a)}$ Below the dotted lines are SGRs. \\
       $^{b)}$ Number of X-ray sources inside the GRB error box \\
       $^{c)}$ (1) Bo\"er \etal (1997),  (2) PI: Bo\"er, 
                  (3) Bo\"er \etal (1994),      (4) PI: Ricker, 
                  (5) McNamara \etal (1995),    (6) Harrison \etal (1996),
                  (7) Hurley \etal (1996),      (8) Marsden \etal (1996),
                  (9) PI: Danner,              (10) Cooke 1993,
                 (11) Greiner (unpubl.) \\
       $^{d)}$ ROSAT detector: P(SPC) or H(RI)
           }
   \vspace{-0.1cm}
   \end{table}

\noindent
However, one has to bear in mind that at the low sensitivity limits reached 
already, the number density of X-ray sources is already remarkably high. From 
the results of many deep pointed observations and combined with the
very deep Lockman hole observations of ROSAT an improved log\,N--log\,S
distribution of X-ray sources has been derived (Hasinger 1997) which
gives 100 (700) X-ray sources per 1\,$\Box$\grad\, 
at the level of 10$^{-14}$ (10$^{-15}$) erg/cm$^2$/s.
Thus, the probability for a chance coincidence of a quiescent, soft X-ray 
source with a GRB location is 25\%--100\% for a 10 arcmin$^2$ size error box.

For the case of GRB error boxes without a detected quiescent X-ray source,
the flux limits of 10$^{-14}$...10$^{-15}$ erg/cm$^2$/s can be used to
constrain particular GRB scenarios (Bo\"er \etal 1994). For instance, 
a neutron star accreting matter from the interstellar material would have
to be at more than $\approx$several kpc distance.

\section{Quick ROSAT follow-up observations of GRBs}

\subsection{Overview}
Motivated by the occurrence of a few  very long lasting GRBs and by the
detection of distinct spectral softening over the burst duration a number of 
attempts have been made to observe well-localized GRB locations with ROSAT as 
quick as possible after the GRB event in the hope to find the ``smoking gun".
To this end, the GRB had to be localized quickly, a target-of-opportunity
observation had to be proposed, and the GRB location had to be within the
ROSAT observing window ($\approx$30\% of the sky at any moment). In several
cases, even a second follow-up observation has been performed to allow
a variability check of the detected X-ray sources. Tab. \ref{toos} lists
the GRBs which have been observed as target-of-opportunity (TOO) together with 
the time delay between the GRB and the ROSAT observation.

   \begin{table}[htb]
     \vspace{-0.1cm}
     \caption{ROSAT TOOs towards GRB locations}
%     \large\hspace*{-1.cm}
     \begin{tabular}{crccl}
      \noalign{\smallskip}
      \hline
      \noalign{\smallskip}
   GRB & $\!\!$Exposure$\!\!$ & Delay & N$_{\rm X}$ & Ref.$^{a)}$ \\
       &  (sec)~~             &       &             &             \\
       \noalign{\smallskip}
      \hline
      \noalign{\smallskip}
   920501       & 2748~~ & 18 days   & 1  & 1 \\
   920711       & 2432~~ & 28 weeks  & 0  & 2, 7 \\
   $\!\!$930704/940301$\!\!$& 3156~~ & ~4 weeks  & 25 & 3, 4 \\
   960720       & 6960~~ & ~6 weeks  & 1  & 5 \\
   960720       & 2791~~ & 24 weeks  &    &  8 \\
   970111       & 1198~~ & ~5 days   & 0  & 6 \\
   970111       &  & ~4 weeks  &    &  8 \\
   970228       &  & 11 days   &    &  8 \\
   \noalign{\smallskip}
   \noalign{$\dashdot$}
   $\!\!$SGR 1806--20 & 1416~~ &  1hr (12 days)$\!\!$ & 0 & 7 \\
      \noalign{\smallskip}
      \hline
      \noalign{\smallskip}
   \end{tabular}
   \label{toos}

   \noindent $^{a)}$ (1) Hurley \etal (1996), 
                     (2) PI: Hurley,          (3) Greiner \etal (1996a),
                     (4) Greiner \etal (1997) (5) Greiner \etal (1996b),
                     (6) Frontera \etal (1997), (7) Greiner (unpubl.), 
                     (8) data not yet available during writing
   \vspace{-0.1cm}
   \end{table}

\begin{figure}[t]
  \centering{
  \hspace{0.01cm}
  \vbox{\psfig{figure=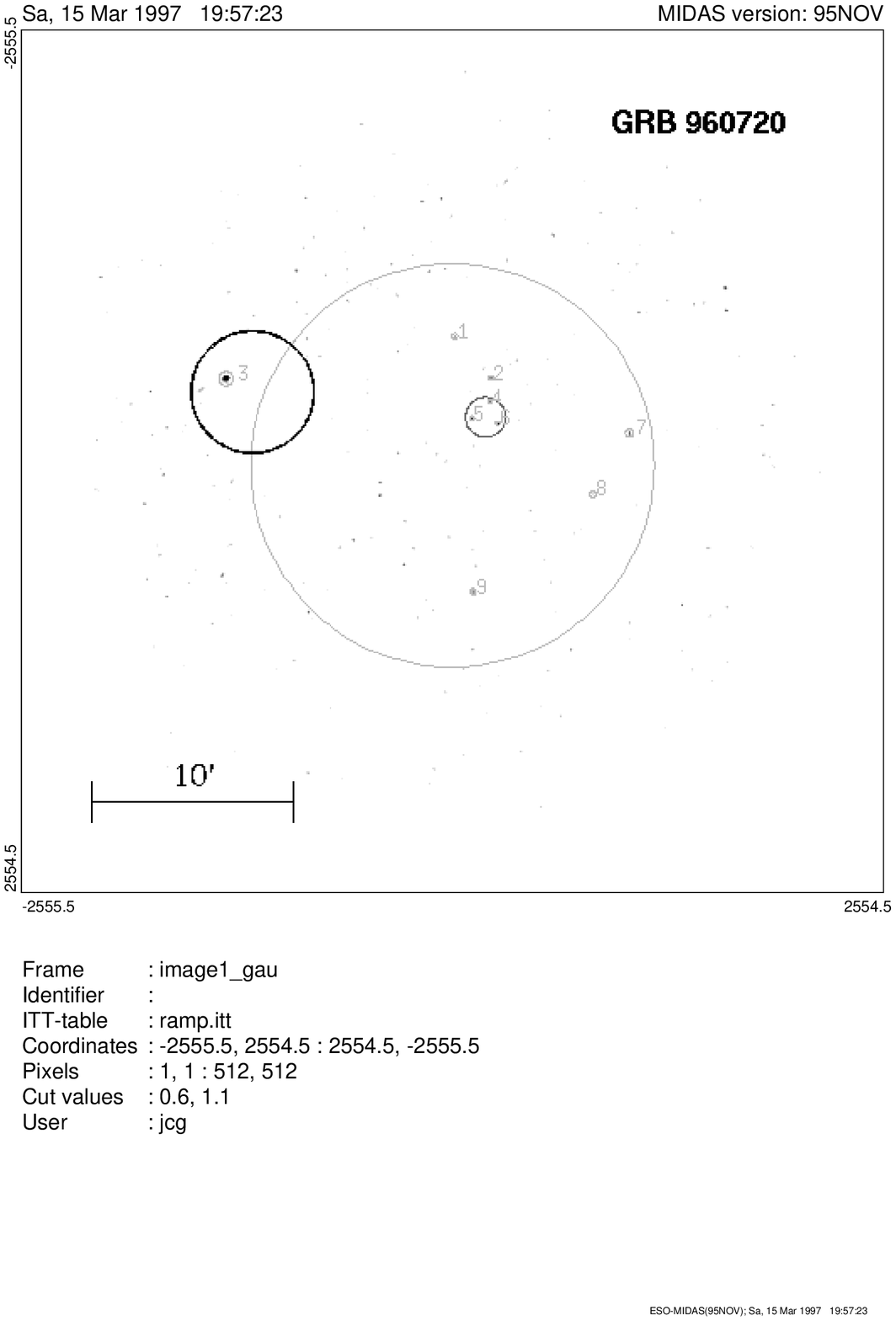,width=8.8cm,%
          bbllx=3.05cm,bblly=10.55cm,bburx=18.7cm,bbury=26.1cm,clip=}}\par
  }
  \caption[grb960720]{Image of the ROSAT HRI observation of GRB 960720.
         The numbered circles are the detected X-ray sources with the size
         of the circles representing the FWHM of the point spread function.
         The bold, 3\amin\, radius circle to the left is the revised GRB error 
         box (Zand \etal 1997), while the large circle is the first reported 
         error box on which the pointing was centered. The small circle
         encompassing the HRI sources \#4, \#5 and \#6 is the steady X-ray 
         source as seen by BeppoSAX and ASCA.
        }
   \label{grb960720}
\end{figure}

The fastest response with ROSAT so far is 5 days which is near the minimum
possible time achievable due to the various scheduling constraints (a curious
exception is the TOO towards SGR 1806--20 which was triggered by one of the 
repeating bursts, and turned out to happen just one hour after another
repeating burst).

An X-ray source has been found in each of the two GRBs 920501 and 960720.
In both cases there is no strong evidence that these X-ray sources may be
related to the GRB event.

\subsection{ROSAT pointings towards the BeppoSAX GRBs} 

The localization of GRBs to an accuracy of several arcmin within less than
a day as achieved for
GRB 960720, 970111 and 970228 by the BeppoSAX wide-field camera (WFC) has 
renewed the interest for really quick follow-up X-ray observations.
While the GRB 960720 location was reported only four weeks after the event,
the ROSAT delay time for TOO pointings has been brought down to 5 days (see
Tab. \ref{toos}). The initial results are shortly summarized below and more
detailed and updated information is accessible over the URL
http:/$\!$/www.aip.de:8080/\~\,jcg/:

GRB 960720 was originally localized with an error circle of 10\amin\, and
observed with ROSAT on August 31, 1996, about 10 days  after the notification
by the BeppoSAX team (Heise, priv. comm.). Nine quiescent X-ray sources were 
detected during the ROSAT HRI pointing, eight of which were within the GRB 
error circle (Greiner \etal 1996b). At the same time, pointings with ASCA and 
the  narrow-field instruments on BeppoSAX  detected one quiescent X-ray source 
(Piro \etal 1996, Murakami \etal 1996) which turned out to be the 
superposition of 3 X-ray sources as seen with the HRI (sources \#4, \#5 and 
\#6 in Fig. \ref{grb960720}). An improvement of the WFC location algorithm
has reduced the error circle of GRB 960720 to 3\amin, and shifted  the
error circle in a way that it contains only one HRI X-ray source (which 
previously was outside the error box). This X-ray source RX J1730.7+4906 
(\#3 in Fig. \ref{grb960720}) has 
been identified with the radio-loud quasar QSO 1729+491 (= 4C 49.29), and 
though the probability of having one X-ray source inside the 3\amin\,
GRB error box is not too small, the probability of having a strong radio 
source in the GRB error circle is as small as 2$\times$10$^{-4}$ (Greiner \&
Heise 1997).

GRB 970111 also was originally localized to 10\amin\, (Costa \etal 1997a).
 Again, a BeppoSAX pointing about 16 hours after the GRB (Butler \etal 1997) 
and a ROSAT HRI observation  5 days after the GRB (Frontera \etal 1997)
revealed several X-ray sources (some of which were also detected during the
ROSAT all-sky survey; see Voges \& Boller 1997)
which later turned out to not be within the revised 3\amin\, GRB error box.

GRB 970228 was localized within a few hours with 3\amin\, accuracy
(Costa \etal 1997b), and  for the first time a fading X-ray source was detected
in BeppoSAX pointings performed 8 hours and 3 days after the GRB
(Costa \etal 1997c). Also a fading radio and optical object have been 
identified within the 50\asec\, error circle of the fading X-ray source.
The ROSAT follow-up observation was possible only 11 days
after the GRB, and the data 
are not yet available during the writing of these lines.

\subsection{X-ray variability time-scale}

For the pointings prior to February 1997 (for which the ROSAT data have been
processed and results are available) the detected X-ray sources have 
(and could only) been checked for possible variability on timescales of months
to years by comparison to both, the ROSAT all-sky survey data (taken 1990/91)
and later follow-up
pointings. Variability on shorter timescales will be checked very soon with 
further pointings on GRB 970111 and 970228.

No fading X-ray source has been identified so far on the timescale of more than
5 days and the sensitivity level as reachable in a few thousend second
ROSAT observation. The recent discovery of a fading X-ray source observed 
8 hours and 3 days after the GRB event by BeppoSAX (Costa \etal 1997c) 
demands to reduce the
ROSAT response time down to 3 days in the future to achieve 10\asec\, size 
error boxes for any upcoming GRB with a fading X-ray source.

\begin{figure*}
  \vbox{\psfig{figure=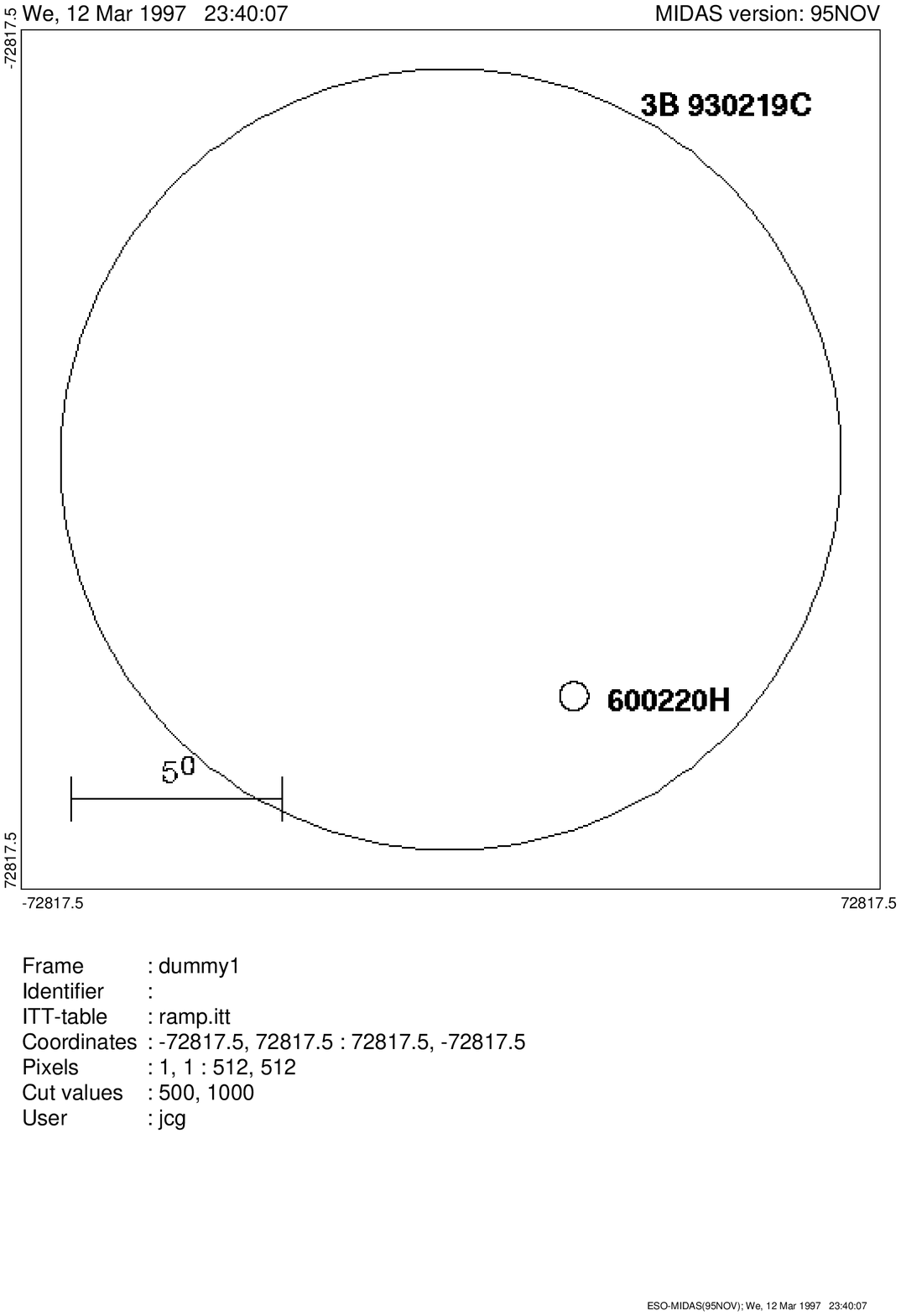,width=8.5cm,%
          bbllx=3.05cm,bblly=10.55cm,bburx=18.7cm,bbury=26.1cm,clip=}}\par
  \vspace*{-8.45cm}\hspace*{9.cm} 
  \vbox{\psfig{figure=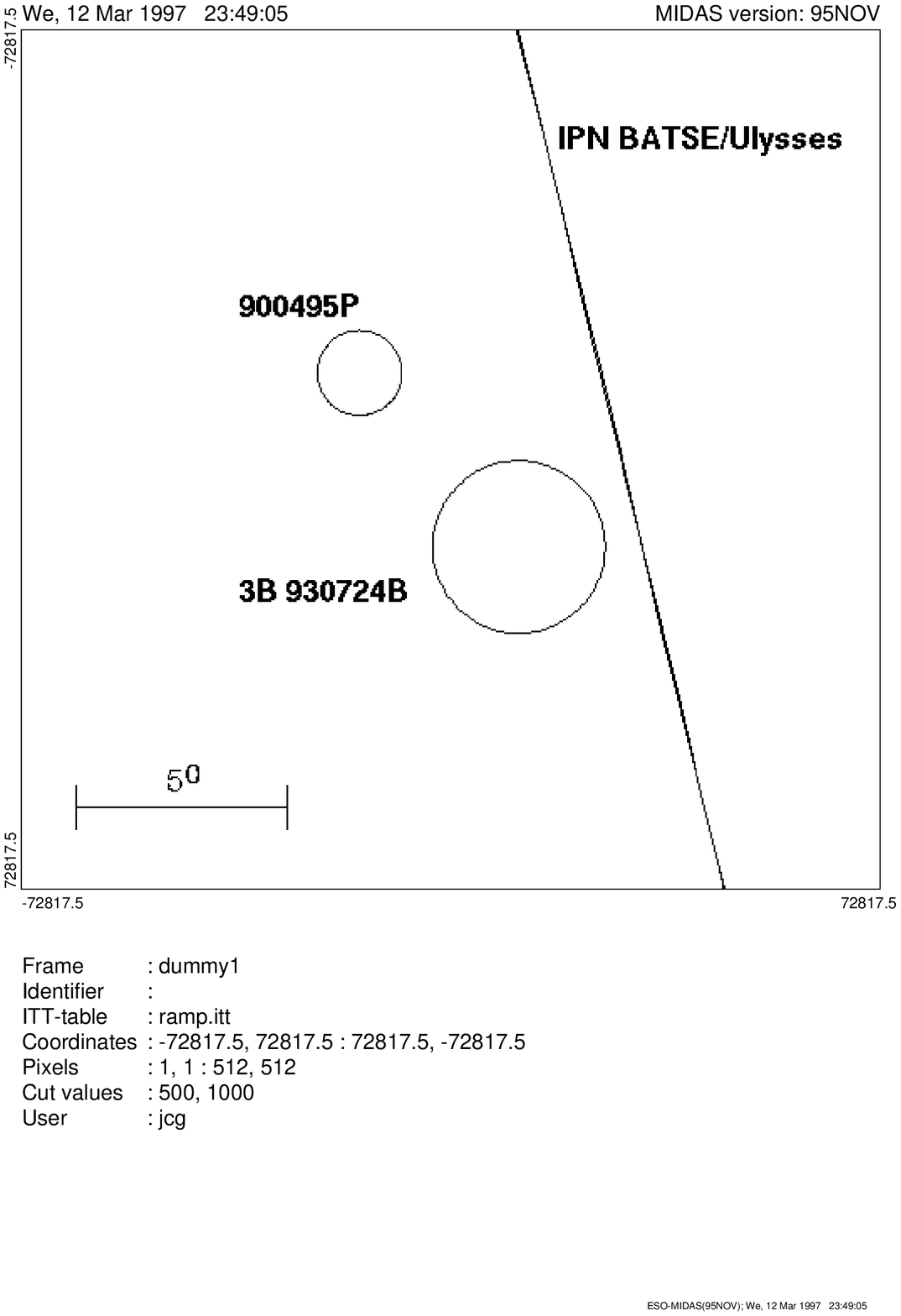,width=8.5cm,%
          bbllx=3.05cm,bblly=10.55cm,bburx=18.7cm,bbury=26.1cm,clip=}}\par
  \caption[simros]{Scetch of the locations of 2 ROSAT pointings simultaneously
       to GRBs: 
       {\bf Left panel:} simultaneous ROSAT HRI observation of 3B 930219C
            covering $\approx$0.15\% of its error box. 
       {\bf Right panel:} simultaneous almost-hit of a ROSAT PSPC 
            observation and 3B 930724B.}
   \label{hits}
%   \vspace*{-.35cm}
\end{figure*}

\section{Correlating GRBs with ROSAT pointings}

Given the $\approx$3 $\Box$\grad\, field of view of the ROSAT PSPC, and
an approximate duty cycle of $\approx$40\% over its 4 yrs livetime, the
chance probability for a serendipitous GRB observation with the ROSAT PSPC
is $\approx$25\%. It is therefore tempting to check whether a GRB occurred
accidentally during a ROSAT observation inside its field of view.

I have correlated the trigger times of the GRBs between April 21, 1991
until Sep. 16, 1994 as  listed in the 3B catalog (Meegan \etal 1996)
with the actual ROSAT observation intervals (OBI) of all pointings
during the given time interval.
If the GRB trigger time is within a ROSAT OBI, then the angular distance is
calculated between the BATSE centroid position of the GRB and the ROSAT
pointing direction. In total, 11(5) cases have been found with a time
coincidence and an angular distance of $<$20\degs(10\degs).

The smallest ever distance is 5\fdg6 for the gamma-ray burst 3B 930724B.
Fortunately, 3B 930724B is quite well localized: it has a 2\fdg0 BATSE error
radius, and an IPN triangulation ring is also available based on the Ulysses
timing (Hurley 1995). Though the IPN arc does not cross the BATSE error
circle, it is clear that the PSPC pointing 900495 is well off the GRB
position (see right panel of Fig. \ref{hits}).
Just to be sure, a timing analysis has been performed on all
sources detected in this pointing, and no transient source is detected.

Among the other cases with less than 20\degs\, angular distance there
exist two hits for which the ROSAT pointing location falls into GRB error
circle: these are the gamma-ray bursts  3B 930219C and 3B 930228.
Of course, these GRB locations are only crudely known, and the ROSAT field
ov view covers only a tiny fraction of the GRB error box. The detailed
numbers for these two GRBs are as follows:
\begin{itemize}
\vspace{-0.1cm} \item 3B 930219C has a 9\fdg2 BATSE error radius, and was not
detected by any other satellite instrument. The ROSAT HRI pointing is 6\fdg2
off the BATSE centroid position, and the HRI field of view covers 0.15\%
of the GRB error box (see left panel of Fig. \ref{hits}). An analysis of
the HRI pointing reveals no transient
of fading X-ray source, thus excluding the occurrence of the GRB within the
location of the HRI pointing.
\vspace{-0.55cm}
\item 3B 930228 has a 18\fdg6 BATSE error radius, and also no
detection by any other instrument. The ROSAT PSPC pointing is 8\fdg9 off
the BATSE centroid position, and the PSPC field of view covers 0.30\%
of the GRB error box. Again, no transient X-ray source is found in the ROSAT
data.
%\vspace{-0.05cm}
\end{itemize}

\section{Search for X-ray flashes in ROSAT data}

Gotthelf \etal (1996) found 42 soft X-ray flashes in a search of 
1.5$\times$10$^7$ s Einstein IPC data. These flashes were argued to be
of astrophysical origin, and are distributed isotropically on the sky.
They do not correlate with any known sources, in particular not with the
nearby galaxy distribution. Based on this result,
Hamilton \etal (1996) conclude that GRBs cannot originate in a galactic halo
with limiting radii between 150--400 kpc.

Vikhlinin (1997) has searched 2.1$\times$10$^7$ s ROSAT PSPC data for the 
occurrence of similar X-ray flashes. Despite greater sensitivity, larger field 
of view and longer exposure as compared to the Einstein data, he does not 
find any flash! This is surprising, because the Einstein X-ray flashes  were 
reported to have soft spectra, so ROSAT should have seen such flashes.
While this leaves the nature of the X-ray flashes detected
with Einstein open for re-evaluation, it invalidates the conclusion regarding
the distance scale of GRBs.

\section{Conclusion}

The search for quiescent X-ray counterparts in small GRB error boxes with ROSAT
is tedious and, in the end, unfortunately, not conclusive so far. This 
strengthens the need for X-ray coverage of future GRB missions. The very 
recent discovery of an X-ray source within the error box of GRB 970228 
fading in intensity over a couple of days  
(Costa \etal 1997c) has opened a new avenue for X-ray follow-up observations
in particular and towards the understanding of GRBs in general.

\vspace{1pc}
\noindent {\it Acknowledgement:}
It is a great pleasure to thank J. Tr\"umper for the steady support of
the GRB research at MPE, and in particular for granting substantial ROSAT time
for target-of-opportunity observations of GRB error boxes.
I'm indebted to J. Englhauser (MPE) for the help in handling the OBI database.
I very much appreciate substantial travel support from the conference 
organizers. 
The author is supported by the German Bundesministerium f\"ur 
Bildung, Wissenschaft, Forschung und Technologie 
(BMBF/DARA) under contract Nos. 50 OR 9201 and 50 QQ 9602 3. 
The ROSAT project is supported by the BMBF/DARA and the Max-Planck-Society.

\section*{References}

\re
Atteia J.-L., Barat C., Hurley K., \etal 1987, ApJ Suppl.  64, 305

\re
Bo\"{e}r M., Atteia J.-L., Gottardi M. \etal 1988, A\&A  202, 117

\re
Bo\"{e}r M., Hurley K., Pizzichini G. \etal 1991, A\&A 249, 118

\re
Bo\"{e}r M., Greiner J., Kahabka P., Motch C., Voges W., Sommer M., Hurley K.,
Niel M., Laros J., Klebesadel R., Kouveliotou C., Fishman G., Cline T.,
1994, 2nd Huntsville workshop 1993, AIP 307, p. 458

\re
Bo\"{e}r M., Motch C., Greiner J., Voges W., Kahabka P., Pedersen H., 1997, 
ApJ (May 20 issue, in press)

\re
Butler R.C., Piro L., Costa E., \etal 1997, IAU Circ. 6539

\re 
Cooke B.A., 1993, Nat. 366, 413

\re
Costa E., Feroci M., Piro L., \etal 1997a, IAU Circ. 6533

\re
Costa E., Feroci M., Frontera F., \etal 1997b, IAU Circ. 6572

\re
Costa E., Feroci M., Piro L., \etal 1997c, IAU Circ. 6576

\re 
Frontera F., Costa E., Piro L., \etal 1997, IAU Circ. 6567

\re
Gotthelf E.V., Hamilton T.T., Helfand D.J., 1996, ApJ 466, 779

\re
Greiner J., Heise J., 1997, IAU Circ. 6570

\re 
Greiner J., Bo\"{e}r M., Motch C., Kahabka P., Voges W.,
1991, Proc. 22nd ICRC Dublin, vol. 1, 53

\re
Greiner J., Bo\"{e}r M., Kahabka P., Motch C., Voges W., 1995a,
NATO ASI C450 ``The Lives of the neutron stars", eds. M.A. Alpar \etal, Kluwer
Acad. Pub., p. 519

\re 
Greiner J., Sommer M., Bade N., Fishman G.J., Hanlon L.O., Hurley K.,
Kippen R.M., Kouveliotou C., Preece R., Ryan J., Sch\"{o}nfelder V.,
Winkler C., O.R. Williams, Bo\"er M., Niel M., 1995b, A\&A 302, 121

\re
Greiner J., Bade N., Hurley K., Kippen R.M., Laros J., 1996a,
3rd Huntsville workshop 1995, eds. C. Kouveliotou \etal, AIP 384, p. 627

\re
Greiner J., Hagen H.-J., Heines A., 1996b, IAU Circ. 6487

\re
Greiner J., Bade N., Hurley K., Kippen R.M., Laros J., 1997, A\&A (in press)

\re
Hamilton T.T., Gotthelf E.V., Helfand D.J., 1996, ApJ 466, 795

\re 
Harrison T.E., McNamara B.J., Williams C.L., Wagner R.M., 1996, AJ 112, 216

\re 
Hasinger G., 1997 (priv. comm.)

\re
Higdon J.C., Lingenfelter R.E., 1990, Ann. Rev. Astron. Astrophys. 28, 401

\re
Hurley K., 1993 (priv. comm.)

\re
Hurley K., 1995, supplemental catalog to the 3B catalog (in electronic form)

\re
Hurley K., Li P., Smette A., Kouveliotou C., Fishman G., Laros J., Cline T., 
Fenimore E., Klebesadel R., Bo\"er M., Pedersen H., Niel M., Sommer M., 1996,
ApJ 464, 342

\re
Lasota J.-P., 1992, in ``Gamma-ray bursts", ed. Ho C. \etal,
Cambridge Univ. Press, p. 17

\re
Marsden D., Rothschild R.E., Lingenfelter R.E., Puetter R.C., 1996, ApJ 470, 513

\re 
McNamara B.J., Harrison T.E., Williams C.L., Wagner R.M., Sokolov V.V., 
Kopylov A.I., Zharykov S.V., 1995, AJ 110, 232

\re
Meegan C.A., Pendleton G.N., Briggs M.S. \etal 1996, ApJ Suppl. 106, 65

\re
Murakami T., Shibata R., Inoue H. \etal 1996, IAU Circ. 6481

\re
Piro L., Costa E., Feroci M. \etal 1996, IAU Circ. 6480

\re
Pizzichini G., Gottardi M., Atteia J.-L. \etal 1986, ApJ 301, 641

\re
Rothschild R., Kulkarni S., Lingenfelter R., 1994, Nat. 368, 432

\re
Tr\"umper J., 1983, Adv. Space Res. 2, 241

\re
Vikhlinin A., 1997 (in prep.)

\re
Voges W., Boller Th., 1997, IAU Circ. 6539

\re
Zand J. in\,'t, Heise J., Hoyng P.,  \etal 1997, IAU Circ. 6569

\label{last}

\end{document}